\documentclass[12pt,english,draftcls, one column]{IEEEtran}
\usepackage[T1]{fontenc}
\usepackage[latin9]{inputenc}
\usepackage{float}
\usepackage{amsmath}
\usepackage{amsthm}
\usepackage{amssymb}
\usepackage{graphicx}
\usepackage{stfloats}
\usepackage{cite} 
\usepackage{xcolor}

\makeatletter

\floatstyle{ruled}
\newfloat{algorithm}{tbp}{loa}
\providecommand{\algorithmname}{Algorithm}
\floatname{algorithm}{\protect\algorithmname}

\theoremstyle{plain}

\theoremstyle{plain}

\ifCLASSINFOpdf
\else
\fi
\usepackage{babel}

\makeatother

\usepackage{babel}
\providecommand{\propositionname}{Proposition}
\providecommand{\theoremname}{Theorem}

\begin{document}

\title{Joint Secure Design of Downlink and D2D Cooperation Strategies for Multi-User Systems}

\author{Seok-Hwan Park, \textit{Member}, \textit{IEEE}, and Xianglan Jin, \textit{Member}, \textit{IEEE}
\thanks{
This work was supported by Basic Science Research Program through the National Research Foundation of Korea (NRF) grants funded by the Ministry of Education [NRF-2019R1A6A1A09031717, 2021R1C1C1006557, 2020R1F1A1068499].
(\textit{Corresponding author: Xianglan Jin}.)

The authors are with the Division of Electronic Engineering and the Future Semiconductor Convergence Technology Research Center, Jeonbuk
National University, Jeonju 54896, Korea (email: seokhwan@jbnu.ac.kr, jinxl77@jbnu.ac.kr).
}}
\maketitle
\begin{abstract}
This work studies the role of inter-user device-to-device (D2D) cooperation for improving physical-layer secret communication in multi-user downlink systems. It is assumed that there are out-of-band D2D channels, on each of which a selected legitimate user transmits an amplified version of the  received downlink signal to other legitimate users. 
A key technical challenge for designing such systems is that eavesdroppers can overhear downlink as well as D2D cooperation signals.
We tackle the problem of jointly optimizing the downlink precoding, artificial noise covariance, and amplification coefficients that maximize  the minimum rate. 
An iterative alternating optimization algorithm is proposed based on the matrix fractional programming. Numerical results confirm the performance gains of the proposed D2D cooperation scheme compared to benchmark secret communication schemes.\end{abstract}


\theoremstyle{theorem}
\newtheorem{theorem}{Theorem}
\theoremstyle{proposition}
\newtheorem{proposition}{Proposition}
\theoremstyle{lemma}
\newtheorem{lemma}{Lemma}
\theoremstyle{corollary}
\newtheorem{corollary}{Corollary}
\theoremstyle{definition}
\newtheorem{definition}{Definition}
\theoremstyle{remark}
\newtheorem{remark}{Remark}

\section{Introduction} \label{sec:intro}

Physical-layer security techniques have been investigated as effective means of improving the communication security of wireless systems \cite{Chu-et-al:TWC16, Park-et-al:SPAWC17, Sun-et-al:TCOM18, Ghanem-et-al:ICC20, Sun-et-al:TIFS20}. Unlike conventional cryptographic techniques which rely on the assumption of limited computational power of eavesdroppers, physical-layer security techniques exploit the physical characteristics of wireless channels to provide perfect communication secrecy.
The authors in \cite{Chu-et-al:TWC16} studied the optimization of transmitter covariance matrix for a single-user multiple-input single-output (MISO) system in the presence of multiple eavesdroppers. 
Secure design for multi-user systems, in which a single base station (BS) communicates with multiple legitimate users, was investigated in \cite{Park-et-al:SPAWC17, Sun-et-al:TCOM18, Ghanem-et-al:ICC20, Sun-et-al:TIFS20}.
These works focus on fronthaul quantization noise design \cite{Park-et-al:SPAWC17}, the application of non-orthogonal multiple access (NOMA) \cite{Sun-et-al:TCOM18}, short-packet transmission for ultra-reliable low-latency communication (URLLC) \cite{Ghanem-et-al:ICC20} and the impact of quantized channel state information (CSI) \cite{Sun-et-al:TIFS20}.

In this work, we study the role of inter-user device-to-device (D2D) cooperation for improving secret communication in multi-user downlink systems. To the best of our knowledge, this is the first work that considers secret communication for D2D-assisted multi-user downlink systems.
The authors in \cite{Karakus-Diggavi:TWC17, Kim-Park:WCL20} investigated the advantages of enabling D2D cooperation assuming pair-wise \cite{Karakus-Diggavi:TWC17} or broadcast cooperation strategy \cite{Kim-Park:WCL20} on out-of-band D2D channels without considering the security issue.
A key technical difficulty in designing D2D cooperation for secret downlink systems is that eavesdroppers have a potential of overhearing both BS-to-users and inter-user D2D  communications. 
This suggests that information leakage by eavesdroppers will reduce the cooperation effect on D2D channels and degrade secrecy rates.

We first describe the system model for a multi-user system with out-of-band D2D communication links (Sec. \ref{sec:system-model}). We then illustrate the signal processing operations at the BS and the selected users for D2D channels including multi-user precoding, artificial noise injection, and amplify-and-forward (AF) D2D relaying (Sec. \ref{sec:secure-downlink-D2D-cooperation}). 
We tackle the problem of jointly optimizing the downlink precoding, artificial noise covariance, and amplification coefficients that maximize the minimum user rate under the constraints on transmit powers and information leakage (Sec. \ref{sec:optimization}). 
To deal with the non-convexity of the original optimization problem, we propose an iterative algorithm based on the matrix fractional programming (FP) \cite{Shen:TN19}.
Numerical results confirm the performance  of the proposed D2D cooperation scheme  compared to conventional secret schemes (Sec. \ref{sec:numerical}).

\section{System Model} \label{sec:system-model}

We consider a multi-user downlink system in which a BS with $M$ transmit antennas serves $K_L$ legitimate single-antenna users. There are $K_E$ single-antenna eavesdropping users that try to overhear the messages intended for legitimate users.
We assume that the $K_L$ users can cooperate one another on $N$ out-of-band D2D cooperation links. 
We define the notations $\mathcal{K}_L\triangleq\{1,2,\ldots,K_L\}$, $\mathcal{K}_E\triangleq\{1,2,\ldots,K_E\}$ and $\mathcal{N}\triangleq\{1,2,\ldots,N\}$.

The received signal of the $k$th legitimate user on the downlink channel is denoted by $y_{L,k,0}$, where the subscripts $L$, $k$, and $0$ stand for `legitimate', the receiving user's index, and the downlink channel, respectively. We write the signal $y_{L,k,0}$ as
\begin{align}
    y_{L,k,0} = \mathbf{h}^{H}_k \mathbf{x}_{B} + z_{L,k,0}, \label{eq:received-signal-legitimate-downlink}
\end{align}
where $\mathbf{x}_{B}\in\mathbb{C}^{M\times 1}$ denotes the transmitted signal vector from the BS, $\mathbf{h}_k \in \mathbb{C}^{M\times 1}$ indicates the channel vector from the BS to the $k$th legitimate user, and $z_{L,k,0}\sim\mathcal{CN}(0,\sigma^2)$ is the additive noise signal.
The transmit signal vector $\mathbf{x}_B$ satisfies the power constraint $\mathtt{E}[||\mathbf{x}_{B}||^2] \leq P_B$.

We assume that, on the $n$th D2D channel with $n\in\mathcal{N}$, the $j_n$th legitimate user transmits a signal $x_{U,j_n}$ to the rest of legitimate users $\mathcal{K}_L\setminus\{j_n\}$. 
The received signal $y_{L,k,n}$ at the $k(\ne j_n)$th legitimate user is then given as
\begin{align}
    y_{L,k,n} = h_{k,j_n}^{\text{d2d}} x_{U,j_n} +  z_{L,k,n}, \label{eq:received-signal-legitimate-D2D}
\end{align}
where $h_{k,j_n}^{\text{d2d}}$ denotes the channel response from the $j_n$th user to the $k$th user, and $z_{L,k,n}\sim\mathcal{CN}(0,\sigma^2)$ represents the additive noise. The transmit signal $x_{U,j_n}$ is subject to the power constraint $\mathtt{E}\left[ |x_{U,j_n}|^2 \right] \leq P_U$.

We note that the  downlink and D2D communication signals can be sensed and overheard by the $K_E$ eavesdroppers.
We denote the signal received by the $m$th eavesdropping user, $m\!\in\!\mathcal{K}_E$, on the downlink channel as $y_{E,m,0}$ which is modelled  by 
\begin{align}
    y_{E,m,0} = \mathbf{g}_m^{H} \mathbf{x}_{B} + z_{E,m,0}, \label{eq:received-signal-eve-downlink}
\end{align}
with $\mathbf{g}_m\in\mathbb{C}^{M\times 1}$ and $z_{E,m,0}\sim\mathcal{CN}(0,\sigma^2)$ representing the channel vector from the BS to the $m$th eavesdropper and the additive noise signal, respectively.
Similarly, the received signal at the $m$th eavesdropper on the $n$th D2D channel, $n\in\mathcal{N}$, is given as
\begin{align}
    y_{E,m,n} = g_{m,j_n}^{\text{d2d}} x_{U,j_n} +  z_{E,m,n}, \label{eq:received-signal-eve-d2d}
\end{align}
where $g_{m,j_n}^{\text{d2d}}$ denotes the channel gain from the $j_n$th legitimate user to the $m$th eavesdropper, and $z_{E,m,n}\sim\mathcal{CN}(0,\sigma^2)$ represents the additive noise. 


\section{Secure Downlink and D2D Cooperation} \label{sec:secure-downlink-D2D-cooperation}

In this section, we describe the signal processing for the secure downlink and D2D cooperation systemS.
On the downlink channel, the BS transmits a superposition of linearly precoded signals and artificial noise, which can be written as $ \mathbf{x}_{B} = \sum\nolimits_{l\in\mathcal{K}_L} \mathbf{v}_{l} s_{l} + \mathbf{n}_{B}$, 
where $s_l\sim\mathcal{CN}(0,1)$ denotes the data signal for the $l$th legitimate user, $\mathbf{v}_{l}\in\mathbb{C}^{M\times 1}$ represents the precoding vector for the signal $s_l$, and $\mathbf{n}_{B}\in \mathbb{C}^{M\times 1}$ is the artificial noise signal injected to prevent the eavesdropping users $\mathcal{K}_E$ from decoding the signals $s_1,s_2,\ldots,s_{K_L}$. Without claim of optimality, we assume that $\mathbf{n}_{B} \sim \mathcal{CN}(\mathbf{0}, \mathbf{Q}_{B})$ with a covariance matrix $\mathbf{Q}_{B}\succeq\mathbf{0}$.
With the transmission model, the power constraint for the BS can be written as
\begin{align}
    \sum\nolimits_{l\in\mathcal{K}_L}||\mathbf{v}_{l}||^2 + \text{tr}\left(\mathbf{Q}_{B}\right) \leq P_B. \label{eq:power-constraint-BS-downlink-precoding-noD2D}
\end{align}
On the $n$th D2D channel, $n\in\mathcal{N}$, the $j_n$th legitimate user broadcasts an amplified version of its received signal $y_{L,j_n,0}$ to help the rest of legitimate users $\mathcal{K}_L\setminus\{j_n\}$ better decode their target signals. The transmitted signal $x_{U,j_n}$ is thus given as $x_{U,j_n} = \alpha_{j_n} y_{L,j_n,0}$.
The amplification coefficient $\alpha_{j_n}$ satisfies the power constraint
\begin{align}
    |\alpha_{j_n}|^2 \cdot p_{r,j_n}\left( \mathbf{v}, \mathbf{Q}_B \right) \leq P_U, \label{eq:power-constraint-user-D2D}
\end{align}
where $p_{r,j_n}( \mathbf{v}, \mathbf{Q}_B ) = \sum_{l\in\mathcal{K}_L} | \mathbf{h}_{j_n}^{H} \mathbf{v}_{l} |^2 + \mathbf{h}_{j_n}^{H}\mathbf{Q}_{B}\mathbf{h}_{j_n} + \sigma^2$ with $\mathbf{v} = \{\mathbf{v}_l\}_{l\in\mathcal{K}_L}$.

The choice of transmitted users $\tilde{\mathcal{K}}_L = \{ j_n \}_{n\in\mathcal{N}}$ on the D2D channels would have a significant impact on the performance. We leave the development of the selection algorithm as a future work and focus on the design of amplification coefficients $\boldsymbol{\alpha} = \{\alpha_{j_n}\}_{n\in\mathcal{N}}$, along with precoding vectors $\mathbf{v}$ and artificial noise covariance $\mathbf{Q}_B$, for fixed set $\tilde{\mathcal{K}}_L$.

The $k$th legitimate user decodes the signal $s_k$ by using the signals $y_{L,k,0}$ and $\{y_{L,k,n}\}_{n\in\mathcal{N}\setminus\{n_k\}}$ received on the downlink and D2D channels, respectively. Here $n_k$ denotes the index of the D2D channel in which the $k$th legitimate user broadcasts an amplification of its received signal. Thus, $n_k$ is an empty element if $k\notin\tilde{\mathcal{K}}_L$.
Combining all the received signals on the downlink and D2D cooperative channels, the achievable data rate $R_k$ for the $k$th legitimate user is given as
\begin{align}
    R_k = I\left(s_k ; y_{L,k,0}, \{y_{L,k,n}\}_{n\in\mathcal{N}\setminus\{n_k\} }\right). \label{eq:secrecy-rate-user-k-withD2D-re}
\end{align}

As in \cite{Park-et-al:TVT18}, we consider information-theoretic privacy constraints:
\begin{align}
    I\left(s_k ; y_{E,m,0}, \{ y_{E,m,n} \}_{n\in\mathcal{N}} \right) \leq \beta, \, m\in\mathcal{K}_E. \label{eq:privacy-constraints}
\end{align}
The left-hand side (LHS) in \eqref{eq:privacy-constraints} measures the amount of information leakage of signal $s_k$ to the $m$th eavesdropper. 
Here we assume that the $m$th eavesdropper can exploit the received signal $y_{E,m,0}$ on the downlink channel, as well as the received signals $\{y_{E,m,n}\}_{n\in\mathcal{N}}$ on the D2D channels. 
Thus, the constraint in (\ref{eq:privacy-constraints}) imposes that the amount of information leakage for every pair of legitimate and eavesdropping users does not exceed a predetermined level $\beta$.
According to information-theoretic results \cite[Ch. 4, Problem 33]{Csiszar-Korner}, a bit stream of rate $R_k^{\text{sec}} = \max(R_k - \beta, 0)$ can be received securely by the $k$th legitimate user  while the remaining rate $\min(\beta, R_k)$ can be overheard by the eavesdropping users. We refer to $R_k^{\text{sec}}$ as the secrecy rate of the $k$th legitimate user.

The respective mutual information  in (\ref{eq:secrecy-rate-user-k-withD2D-re}) and (\ref{eq:privacy-constraints}) can be written as
\begin{align}
    &I\left(s_k ; y_{L,k,0}, \{y_{L,k,n}\}_{n\in\mathcal{N}\setminus\{n_k\} }\right)   \nonumber \\ 
    &\!\!\!\!= f_{L,k}\left(\mathbf{v}, \mathbf{Q}_B, \boldsymbol{\alpha}\right) 
    =\log_2\left(1 + \mathbf{v}_k^H \bar{\mathbf{H}}_k^H \mathbf{C}_{L,k}^{-1}\bar{\mathbf{H}}_k\mathbf{v}_k \right),  \label{eq:MI-computation-data-rate} \\
    &I\left(s_k ; y_{E,m,0}, \{ y_{E,m,n} \}_{n\in\mathcal{N}} \right) \nonumber \\ 
    &\!\!\!\!\!=\! f_{E,k,m}\left(\mathbf{v}, \mathbf{Q}_B, \boldsymbol{\alpha}\right)
    \!=\! \log_2\left(\! 1 \!+\! \mathbf{v}_k^H\bar{\mathbf{G}}_{m}^H \mathbf{C}_{E,k,m}^{-1} \bar{\mathbf{G}}_{m}\mathbf{v}_k \right)\!,  \label{eq:MI-computation-privacy}
\end{align}
where we have defined 
\begin{align*}
\mathbf{C}_{L,k} &=  \bar{\mathbf{H}}_k^{} \mathbf{Q}_B  \bar{\mathbf{H}}_k^{H} \!+\! \sigma^2 \mathbf{I} + \bar{\mathbf{Z}}_{L,k} \!+\! \sum_{l\in\mathcal{K}_L\setminus\{k\}} \bar{\mathbf{H}}_k^{}\mathbf{v}_l\mathbf{v}_l^H\bar{\mathbf{H}}_k^{H},\\
\mathbf{C}_{E,k,m} \!&= \bar{\mathbf{G}}_{m}\mathbf{Q}_B\bar{\mathbf{G}}_{m}^H \!+\! \sigma^2 \mathbf{I} \!+\! \bar{\mathbf{Z}}_{E,m} \!+\!\! \sum_{l\in\mathcal{K}_L\!\setminus\{k\}} \bar{\mathbf{G}}_{m}\mathbf{v}_l\mathbf{v}_l^H\bar{\mathbf{G}}_{m}^H,
\end{align*}
and $\boldsymbol{\alpha} = \{\alpha_{j_n}\}_{n\in\mathcal{N}}$. The matrices $\bar{\mathbf{H}}_k \in \mathbb{C}^{(N+1)\times M}$, $\bar{\mathbf{G}}_{m} \in \mathbb{C}^{(N+1)\times M}$,
$\bar{\mathbf{Z}}_{L,k} \in \mathbb{C}^{(N+1) \times (N+1)}$ and $\bar{\mathbf{Z}}_{E,m} \in \mathbb{C}^{(N+1)\times (N+1)}$ are defined as
\begin{align}
    \bar{\mathbf{H}}_k &= \left[ \mathbf{h}_k^{H} ;\,\, \tilde{\mathbf{h}}_{k,1}^{H} ;\, \cdots \,; \tilde{\mathbf{h}}_{k,N}^{H} \right], \nonumber \\
    \bar{\mathbf{G}}_{m} &= \left[ \mathbf{g}_m^{H} ; \,\, g_{m,k_1}^{\text{d2d}}\alpha_{j_1}\mathbf{h}_{j_1}^{H} ;\, \cdots ;\, g_{m,j_N}^{\text{d2d}}\alpha_{j_N}\mathbf{h}_{j_N}^{H} \right], \nonumber \\
    \bar{\mathbf{Z}}_{L,k} &= \text{diag}\left(0, \tilde{\alpha}_{k,j_1}, \ldots, \tilde{\alpha}_{k,j_N}\right), \nonumber \\
    \bar{\mathbf{Z}}_{E,m} & = \text{diag}\left( 0, \{\sigma^2 |g_{m,j_n}^{\text{d2d}}|^2|\alpha_{j_n}|^2\}_{n\in\mathcal{N}} \right), \nonumber
\end{align}
with $\tilde{\mathbf{h}}_{k,n}^{H} = \mathtt{1}_{j_n \neq k} \cdot h_{k,j_n}^{\text{d2d}}\alpha_{j_n}\mathbf{h}_{j_n}^{H}$ and $\tilde{\alpha}_{k,n} = \mathtt{1}_{j_n \neq k} \cdot\sigma^2 |h_{k,j_n}^{\text{d2d}}|^2|\alpha_{j_n}|^2$. Here $\mathtt{1}_{(\cdot)}$ denotes an indicator function, which returns 1 if the statement in the subscript is true or 0 otherwise.

\section{Proposed Optimization} \label{sec:optimization}

The goal is to jointly optimize the precoding vectors $\mathbf{v}$, the artificial noise covariance matrix $\mathbf{Q}_{B}$, and the amplification coefficients $\boldsymbol{\alpha}$ with the criterion of maximizing the minimum-user rate $R_{\min} = \min_{k\in\mathcal{K}_L} R_k$ while satisfying the transmit power constraints in  (\ref{eq:power-constraint-BS-downlink-precoding-noD2D}) and (\ref{eq:power-constraint-user-D2D}), and the privacy constraints in (\ref{eq:privacy-constraints}). 
Let $\mathbf{R}=[R_1;\cdots;R_L]$.
We formulate the problem at hand as
\begin{subequations} \label{eq:problem-D2D-original}
\begin{align}
    &\underset{\mathbf{R},\mathbf{v}, \mathbf{Q}_{B}\succeq \mathbf{0}, \boldsymbol{\alpha}}{\mathrm{maximize}}\,\,  R_{\min} \label{eq:problem-D2D-original-objective}\\
    &\qquad\mathrm{s.t.}\,\,\,  R_k \leq f_{L,k}\left(\mathbf{v}, \mathbf{Q}_B, \boldsymbol{\alpha}\right), ~~k\in\mathcal{K}_L, \label{eq:problem-D2D-original-rate} \\
    & \qquad\qquad f_{E,k, m}\left(\mathbf{v}, \mathbf{Q}_B, \boldsymbol{\alpha}\right) \leq \beta, ~k\in\mathcal{K}_L, m\in\mathcal{K}_E, \label{eq:problem-D2D-original-privacy} \\
    & \qquad\qquad\sum\nolimits_{l\in\mathcal{K}_L}||\mathbf{v}_{l}||^2 + \text{tr}\left(\mathbf{Q}_{B}\right) \leq P_B, \label{eq:problem-D2D-original-power-BS} \\
    & \qquad\qquad|\alpha_{j_n}|^2 \cdot p_{r,j_n}\left( \mathbf{v}, \mathbf{Q}_B \right) \leq P_U, ~n\in\mathcal{N}. \label{eq:problem-D2D-original-power-UE}
\end{align}
\end{subequations}

It is difficult to solve the problem (\ref{eq:problem-D2D-original}) due to the non-convexity of the constraints (\ref{eq:problem-D2D-original-rate}), (\ref{eq:problem-D2D-original-privacy}) and (\ref{eq:problem-D2D-original-power-UE}). 
To handle this issue, we restate the problem by using an epigraph form \cite[Ch. 4.1.3]{Convex} and the matrix fractional programming (FP) \cite{Shen:TN19}, and propose an alternating optimization approach.


%

Applying the FP \cite[Cor. 1]{Shen:TN19},  (\ref{eq:problem-D2D-original-rate}) can be replaced by a stricter constraint: 
\begin{align}
    \!R_k \!\leq\! \phi\left( \gamma_{k, \text{info}} \boldsymbol{,}\, \boldsymbol{\theta}_{k,\text{info}}\boldsymbol{,}\, \bar{\mathbf{H}}_k^{}\mathbf{v}_k\boldsymbol{,}\, \mathbf{C}_{L,k} \!+\! \bar{\mathbf{H}}_k \mathbf{v}_k \mathbf{v}_k^H \bar{\mathbf{H}}_k^H \right), \label{eq:rewrite-rate-constraint-info-FP}
\end{align}
 for any $\gamma_{k, \text{info}}$ and $\boldsymbol{\theta}_{k,\text{info}} \in \mathbb{C}^{(N+1)\times 1}$, $ k\in\mathcal{K}_L$,
 where $\phi (\mathbf{A}, \mathbf{B}, \mathbf{C}, \mathbf{D}) = \log_2\det(\mathbf{I} + \mathbf{A}) - \text{tr}(\mathbf{A})/\ln 2 + \text{tr}( (\mathbf{I}+\mathbf{A})(2\mathbf{C}^H\mathbf{B} - \mathbf{B}^H\mathbf{D}\mathbf{B}) )/\ln 2$.
Note that the constraint in (\ref{eq:rewrite-rate-constraint-info-FP}) is biconvex with respect to $\{\mathbf{v}, \tilde{\mathbf{Q}}_B\}$ and $\boldsymbol{\alpha}$, if the auxiliary variables $\gamma_{k, \text{info}}$ and $\boldsymbol{\theta}_{k,\text{info}}$ are fixed, where $\tilde{\mathbf{Q}}_B = \mathbf{Q}_B^{1/2}$. Also, (\ref{eq:rewrite-rate-constraint-info-FP}) becomes equivalent to (\ref{eq:problem-D2D-original-rate}) when $\gamma_{k, \text{info}}$ and $\boldsymbol{\theta}_{k,\text{info}}$ are given as
\begin{subequations} \label{eq:update-auxiliary-info-D2D}
\begin{align}
    \gamma_{k,\text{info}} &= \mathbf{v}_k^H\bar{\mathbf{H}}_k^H \mathbf{C}_{L,k}^{-1}  \bar{\mathbf{H}}_k\mathbf{v}_k, \label{eq:update-gamma-info-D2D} \\
    \boldsymbol{\theta}_{k,\text{info}} &= \left(\mathbf{C}_{L,k} + \bar{\mathbf{H}}_k\mathbf{v}_k\mathbf{v}_k^H\bar{\mathbf{H}}_k^H \right)^{-1} \bar{\mathbf{H}}_k \mathbf{v}_k. \label{eq:update-theta-info-D2D}
\end{align}
\end{subequations}
Although (\ref{eq:problem-D2D-original-rate}) and (\ref{eq:problem-D2D-original-privacy}) are of similar forms, we cannot directly apply the FP for (\ref{eq:problem-D2D-original-privacy}).
We first restate (\ref{eq:problem-D2D-original-privacy}) as
\begin{subequations} \label{eq:epigraph-leak-D2D}
\begin{align}
    \beta~  &\geq A_{m,\text{leak}} - B_{k,m,\text{leak}}, \label{eq:epigraph-leak-D2D-1} \\
    A_{m,\text{leak}} &\geq \log_2\det\left( \mathbf{C}_{E,k,m} + \bar{\mathbf{G}}_m \mathbf{v}_k\mathbf{v}_k^H\bar{\mathbf{G}}_m^H \right),\label{eq:epigraph-leak-D2D-2} \\
    B_{k,m,\text{leak}} &\leq \log_2\det\left( \mathbf{C}_{E,k,m}\right), \label{eq:epigraph-leak-D2D-3}
\end{align}
\end{subequations}
for $k\in\mathcal{K}_L, m\in\mathcal{K}_E$.
The constraint in (\ref{eq:epigraph-leak-D2D-1}) is convex since both sides are affine functions.
The constraint in (\ref{eq:epigraph-leak-D2D-3}) has a similar form to (\ref{eq:problem-D2D-original-rate}), and we can use FP to obtain a stricter condition \cite[Cor. 1]{Shen:TN19}:
\begin{align}
    B_{k,m,\text{leak}} &\leq \phi\left( \boldsymbol{\Gamma}_{k,m,\text{leak}}\boldsymbol{,} \boldsymbol{\Theta}_{k,m,\text{leak}}\boldsymbol{,} \hat{\mathbf{G}}_m \hat{\mathbf{V}}_{k,m}\boldsymbol{,} \hat{\mathbf{C}}_{E,k,m}\right) \nonumber \\
    & \qquad+ (N+1)\log_2\sigma^2, \,\, k\in\mathcal{K}_L, m\in\mathcal{K}_E, \label{eq:rewrite-FP-epigraph-leak-D2D-3}
\end{align}
for any $\boldsymbol{\Gamma}_{k,m,\text{leak}} \in \mathbb{C}^{(M+K_L+N)\times(M+K_L+N)}$ and $\boldsymbol{\Theta}_{k,m,\text{leak}} \in \mathbb{C}^{(N+1) \times (M+K_L+N)}$.
Here the matrices $\hat{\mathbf{C}}_{E,k,m}\in\mathbb{C}^{(N+1)\times(N+1)}$, $\hat{\mathbf{G}}_m \in \mathbb{C}^{(N+1)\times (MK_L + N+1)}$ and $\hat{\mathbf{V}}_{k,m} \in \mathbb{C}^{ (M K_L + N+1) \times (M+K_L+N) }$ are defined as
\begin{align}
    \hat{\mathbf{C}}_{E,k,m} &= \sigma^2\mathbf{I} + \hat{\mathbf{G}}_m \hat{\mathbf{V}}_{k,m} \hat{\mathbf{V}}_{k,m}^H \hat{\mathbf{G}}_m^H, \nonumber \\
    \hat{\mathbf{G}}_m &= \left[ \left(\mathbf{1}_{1\times K_L} \otimes \bar{\mathbf{G}}_m\right) \,\,\, \mathbf{I}_{N+1} \right], \nonumber\\
    \hat{\mathbf{V}}_{k,m} & = \text{diag} \left( \tilde{\mathbf{Q}}_B, \{\mathbf{v}_l\}_{l\in\mathcal{K}_L\setminus\{k\}}, \tilde{\mathbf{Z}}_{E,m} \right), \nonumber
\end{align}
with $\tilde{\mathbf{Z}}_{E,m} = \mathbf{Z}_{E,m}^{1/2} \in \mathbb{C}^{(N+1)\times(N+1)}$.
Similar to (\ref{eq:rewrite-rate-constraint-info-FP}), the constraint (\ref{eq:rewrite-FP-epigraph-leak-D2D-3}) is biconvex with respect to $\{\mathbf{v},\tilde{\mathbf{Q}}_B\}$ and $\boldsymbol{\alpha}$ for fixed $\boldsymbol{\Gamma}_{k,m,\text{leak}}$ and $\boldsymbol{\Theta}_{k,m,\text{leak}}$,
and it is equivalent to (\ref{eq:epigraph-leak-D2D-3}) if
\begin{subequations} \label{eq:update-auxiliary-leak-B-D2D}
\begin{align}
    \boldsymbol{\Gamma}_{k,m,\text{leak}} &= \frac{1}{\sigma^2} \hat{\mathbf{V}}_{k,m}^H \hat{\mathbf{G}}_m^H \hat{\mathbf{G}}_m \hat{\mathbf{V}}_{k,m}, \label{eq:update-Gamma-leak-D2D}\\
    \boldsymbol{\Theta}_{k,m,\text{leak}} &= \hat{\mathbf{C}}_{E,k,m}^{\,\,-1} \hat{\mathbf{G}}_m \hat{\mathbf{V}}_{k,m}. \label{eq:update-Theta-leak-D2D}
\end{align}
\end{subequations}
Lastly, we discuss how to handle the non-convex constraint (\ref{eq:epigraph-leak-D2D-2}) which has the opposite inequality direction to (\ref{eq:problem-D2D-original-rate}).
By using the result in \cite[Lem. 1]{Zhou-Yu:TSP16}, we can see that (\ref{eq:epigraph-leak-D2D-2}) is satisfied, if there exists $\mathbf{\Sigma}_{m,\text{leak}} \in \mathbb{C}^{(N+1)\times(N+1)}$ that satisfies
\begin{align}
    A_{m,\text{leak}} &\geq \log_2\det\left(\mathbf{\Sigma}_{m,\text{leak}}\right) - (N+1)/\ln 2 \label{eq:rewrite-epigraph-leak-D2D-2} \\
    & + \frac{1}{\ln 2} \text{tr}\left( \mathbf{\Sigma}_{m,\text{leak}}^{-1} \left( \mathbf{C}_{E,k,m} + \bar{\mathbf{G}}_m\mathbf{v}_k\mathbf{v}_k^H\bar{\mathbf{G}}_m^H \right) \right), \, m\in\mathcal{K}_E. \nonumber 
\end{align}
It is straightforward to see that the above constraint is biconvex with respect to $\{\mathbf{v},\mathbf{Q}_B\}$ and $\boldsymbol{\alpha}$ for fixed $\mathbf{\Sigma}_{m,\text{leak}}$, and that it is equivalent to (\ref{eq:epigraph-leak-D2D-2}) if we have
\begin{align}
    \!\!\mathbf{\Sigma}_{m,\text{leak}} = \mathbf{C}_{E,k,m} + \bar{\mathbf{G}}_m\mathbf{v}_k\mathbf{v}_k^H\bar{\mathbf{G}}_m^H. \label{eq:update-Sigma-leak-D2D}
\end{align}

Based on the above observations, we consider the following problem which has the same optimal solution as in  (\ref{eq:problem-D2D-original}):
\begin{align}
    &\underset{^{\mathbf{R},\mathbf{v}, \tilde{\mathbf{Q}}_{B}, \boldsymbol{\alpha}, \boldsymbol{\gamma}, \boldsymbol{\theta},} _{\,\,\,\boldsymbol{\Gamma}, \boldsymbol{\Theta}, \boldsymbol{\Sigma}, \mathbf{A}, \mathbf{B}}}{\mathrm{maximize}}\,\,  R_{\min} \label{eq:problem-D2D-restated} \\
    &\qquad\mathrm{s.t.}\,\,\,  R_k: \text{(\ref{eq:rewrite-rate-constraint-info-FP})}, \nonumber \\
    &\qquad\qquad \beta \geq A_{m,\text{leak}} - B_{k,m,\text{leak}}, k\in\mathcal{K}_L, m\in\mathcal{K}_E, \nonumber \\
    &\qquad\qquad A_{m,\text{leak}}: \text{(\ref{eq:rewrite-epigraph-leak-D2D-2})}, B_{k,m,\text{leak}}: \text{(\ref{eq:rewrite-FP-epigraph-leak-D2D-3})}, \nonumber \\ 
    &\qquad\qquad \sum\nolimits_{l\in\mathcal{K}_L}||\mathbf{v}_{l}||^2 + \text{tr}\left(\mathbf{Q}_{B}\right) \leq P_B, \nonumber \\
    &\qquad\qquad |\alpha_{j_n}|^2 \cdot p_{r,j_n}\left( \mathbf{v}, \mathbf{Q}_B \right) \leq P_U, n\in\mathcal{N}. \nonumber
\end{align}

The restated problem (\ref{eq:problem-D2D-restated}) is biconvex with respect to $\{\mathbf{v},\tilde{\mathbf{Q}}_B\}$ and $\boldsymbol{\alpha}$, if the auxiliary variables $\{\boldsymbol{\gamma}, \boldsymbol{\Gamma}, \boldsymbol{\theta}, \boldsymbol{\Theta}, \boldsymbol{\Sigma}\}$ are fixed.
Also, the optimal auxiliary variables $\{\boldsymbol{\gamma}, \boldsymbol{\Gamma}, \boldsymbol{\theta}, \boldsymbol{\Theta}, \boldsymbol{\Sigma}\}$ can be obtained in closed-form expressions (\ref{eq:update-auxiliary-info-D2D}), (\ref{eq:update-auxiliary-leak-B-D2D}), and (\ref{eq:update-Sigma-leak-D2D}).
Inspired by this observation, we propose an iterative algorithm which alternately updates the variable sets $\{\mathbf{v}, \tilde{\mathbf{Q}}_{B}\}$, $\boldsymbol{\alpha}$, and $\{\boldsymbol{\gamma}, \boldsymbol{\Gamma}, \boldsymbol{\theta}, \boldsymbol{\Theta}, \boldsymbol{\Sigma}\}$. 
The detailed algorithm is described in Algorithm 1. 
We note that the objective minimum rate $R_{\min}^{(t)}$ monotonically increases with the iteration index $t$ so that it converges to a locally optimal point of problem (\ref{eq:problem-D2D-restated}).

\begin{algorithm}
\caption{Proposed alternating optimization algorithm}

\textbf{1.} Initialize $\{\mathbf{v}, \tilde{\mathbf{Q}}_{B}, \boldsymbol{\alpha}\}$ as arbitrary values that satisfy the power constraints (\ref{eq:problem-D2D-original-power-BS}) and (\ref{eq:problem-D2D-original-power-UE}), and set $t\leftarrow 1$.

\textbf{2.} Compute the minimum rate $R_{\min}$ with the initialized $\{\mathbf{v}, \tilde{\mathbf{Q}}_{B}, \boldsymbol{\alpha}\}$, and set $R_{\min}^{(0)} \leftarrow R_{\min}$.

\textbf{3.} Update $\{\boldsymbol{\gamma}, \boldsymbol{\Gamma}, \boldsymbol{\theta}, \boldsymbol{\Theta}, \boldsymbol{\Sigma}\}$ according to (\ref{eq:update-auxiliary-info-D2D}), (\ref{eq:update-auxiliary-leak-B-D2D}) and (\ref{eq:update-Sigma-leak-D2D}).

\textbf{4.} Update $\{\mathbf{v}, \tilde{\mathbf{Q}}_{B}\}$ as a solution to the convex problem obtained by fixing $\boldsymbol{\alpha}$ and $\{\boldsymbol{\gamma}, \boldsymbol{\Gamma}, \boldsymbol{\theta}, \boldsymbol{\Theta}, \boldsymbol{\Sigma}\}$ in problem (\ref{eq:problem-D2D-restated}).

\textbf{5.} Repeat Step 3.

\textbf{6.} Update $\boldsymbol{\alpha}$ as a solution to the convex problem obtained by fixing $\{\mathbf{v}, \tilde{\mathbf{Q}}_{B}\}$ and $\{\boldsymbol{\gamma}, \boldsymbol{\Gamma}, \boldsymbol{\theta}, \boldsymbol{\Theta}, \boldsymbol{\Sigma}\}$ in problem (\ref{eq:problem-D2D-restated}).

\textbf{7.} Compute the minimum rate $R_{\min}$ with the updated $\{\mathbf{v}, \tilde{\mathbf{Q}}_{B}, \boldsymbol{\alpha}\}$, and set $R_{\min}^{(t)} \leftarrow R_{\min}$.

\textbf{8.} If $|R_{\min}^{(t)} - R_{\min}^{(t-1)}| \leq \delta$ or $t > t_{\max}$, stop. Otherwise, go back to Step 3 with setting $t\leftarrow t+1$.

\end{algorithm}

The complexity of Algorithm 1 is depending on the computational complexity in each iteration and the number of iterations. The former is dominated by solving the convex problems in Steps 4 and 6. It was shown in \cite[p. 4]{BTal-Nemirovski} that for a given error tolerance level $\epsilon$, the complexity of solving a convex problem is upper bounded by $\mathcal{O}( n_O(n_O^3+n_A) \log(1/\epsilon))$, where $n_O$ and $n_A$ respectively represent the numbers of optimization variables and the arithmetic operations needed to compute the objective and constraint functions. 
The numbers $n_O$ and $n_A$ for the convex problem tackled at Step 4 are given as $n_O = 2MK_{L}+M^{2}+2K_{L}+K_{E}+K_{L}K_{E}$ and $n_A = (2M+7N)NK_{L}+4(2M^{2}+K_{L}(M+N)+N^{2})NK_{E}+4(M+K_{L}+N)(N(MK_{L}+N)+(M+K_{L}+N)(M+K_{L}+4N))K_{L}K_{E}+4(K_{L}+2M)MN$, and those for Step 6 are given as $n_O = 2N+2K_{L}+K_{E}+K_{L}K_{E}$ and $n_A = (2M+7N)NK_{L}+4(2M^{2}+K_{L}(M+N)+N^{2})NK_{E}+4(M+K_{L}+N)(N(MK_{L}+N)+(M+K_{L}+N)(M+K_{L}+4N))K_{L}K_{E}$.
Moreover, numerical checks show that Algorithm 1 converges within a few tens of iterations.

\section{Numerical Results} \label{sec:numerical}

We assume that the BS is located at the center $(0,0)$ of a rectangular area of side length 100, and the $K_L$ legitimate users are randomly located within the area. The $K_E$ eavesdropping users are randomly located in another rectangular area of the same shape, but with different center point $(100,0)$. 
We adopt the path-loss model $c_0(d/d_0)^{-\eta}$ \cite{Kim-Park:WCL20}, where $d$ denotes the distance between the transmitting and receiving nodes, and $c_0$ and $d_0$ are set to $10$ dB and $30$, respectively.
The $N$ transmitted users in $\tilde{\mathcal{K}}_L$ are randomly chosen from $\mathcal{K}_L$.
We also assume independent and identically distributed (i.i.d.) Rayleigh small-scale fading model. 
We compare the average performance over various channel realizations for the following schemes: \textit{i)} Proposed D2D: $\{\mathbf{v},\mathbf{Q}_B,\boldsymbol{\alpha}\}$ are jointly optimized according to Algorithm 1; \textit{ii)} No D2D: Algorithm 1 is executed while fixing $\alpha_{j_n}=0$ for all $n\in\mathcal{N}$ and skipping Steps 4 and 5; \textit{iii)} Random D2D: Each $\alpha_{j_n}$ is fixed as $\alpha_{j_n}\leftarrow (\tilde{\alpha}_{j_n} / |\tilde{\alpha}_{j_n}|) (P_U/p_{r,j_n} (\mathbf{v},\mathbf{Q}_B))^{1/2} $, where $\tilde{\alpha}_{j_n}$ follows $\tilde{\alpha}_{j_n}\sim \mathcal{CN}(0,1)$, and the received power $p_{r,j_n} (\mathbf{v},\mathbf{Q}_B)$ is computed using $\{\mathbf{v},\mathbf{Q}_B\}$ of no D2D scheme.
Then, $\{\mathbf{v},\mathbf{Q}_B\}$ are optimized for fixed $\boldsymbol{\alpha}$ by applying Steps 2, 3, and 6.
Note that the baseline schemes require lower complexity than the proposed scheme since the coefficients $\boldsymbol{\alpha}$ are fixed at the sacrifice of performance.

\begin{figure}
\centering\includegraphics[width=11cm,height=9.0cm,keepaspectratio]{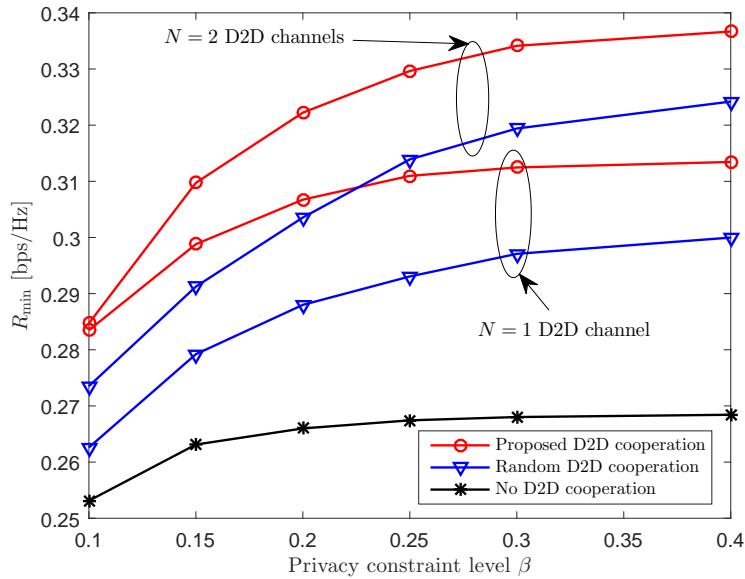}\caption{{\scriptsize\label{fig:graph-D2D-vs-noD2D-wrt-interAreaDist}{}$R_{\min}$ versus
the privacy constraint level $\beta$ for $M=2$, $K_L=8$, $K_E=2$, $N\in\{1,2\}$ and $P_B/\sigma^2 = P_U/\sigma^2 = 10$ dB}}
\end{figure}

In Fig. \ref{fig:graph-D2D-vs-noD2D-wrt-interAreaDist}, we plot  $R_{\min}$ by increasing the privacy constraint level $\beta$ for a multi-user system with $M=2$, $K_L=8$, $K_E=2$, $N\in\{1,2\}$ and $P_B/\sigma^2 = P_U/\sigma^2 = 10$ dB.
The figure shows that $R_{\min}$ is significantly improved by D2D cooperation either with optimized $\boldsymbol{\alpha}$ or with random $\boldsymbol{\alpha}$.
Furthermore, we observe that $R_{\min}$ increases with $\beta$. This is consistent with the fact that the maximized objective value improves when the constraint becomes looser (see \eqref{eq:problem-D2D-original}).

\begin{figure}
\centering\includegraphics[width=11cm,height=9.0cm,keepaspectratio]{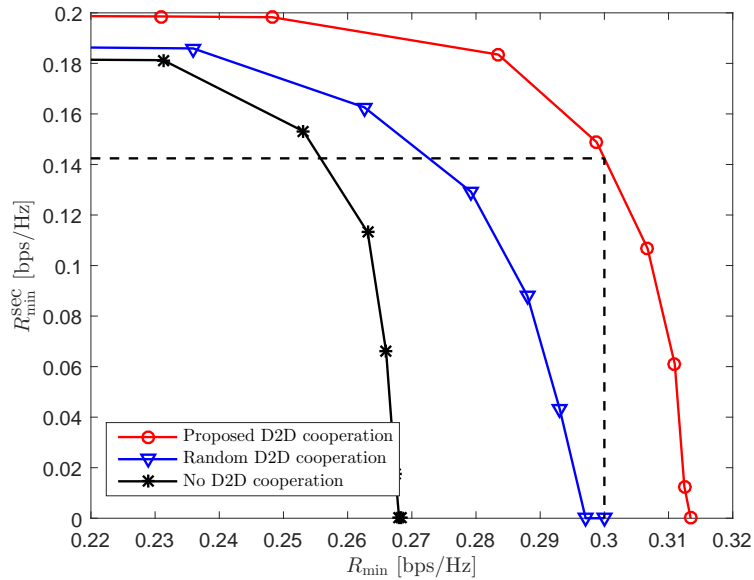}\caption{{\scriptsize\label{fig:graph-tradeOff}$R_{\min}^{\text{sec}}$ versus
$R_{\min}$ for $M=2$, $K_L=8$, $K_E=2$, $N=1$ and $P_B/\sigma^2 = P_U/\sigma^2=10$ dB}}
\end{figure}

In Fig. \ref{fig:graph-tradeOff}, we plot the minimum secrecy rate $R_{\min}^{\text{sec}}=\min_{k\in\mathcal{K}_L} R_k^{\text{sec}}$ versus the minimum rate $R_{\min}$ for a multi-user system with $M=2$, $K_L=8$, $K_E=2$, $N=1$ and $P_B/\sigma^2 = P_U/\sigma^2=10$ dB.
Different points of each scheme are obtained for different privacy levels $\beta$. That is, with a larger $\beta$, we achieve larger $R_{\min}$ but less $R_{\min}^{\text{sec}}$. 
It is also worth noting that in order to achieve $R_{\min} = 0.3$ bps/Hz, the minimum secrecy rates of random D2D and no D2D schemes are degraded to $R_{\min}^{\text{sec}} = 0$ while the proposed D2D scheme achieves $R_{\min}^{\text{sec}} \geq 0.14$ bps/Hz.

\section{Conclusion}

We have studied the advantages of enabling D2D cooperation for secret multi-user downlink systems in the presence of eavesdroppers, which can overhear both the downlink and D2D cooperation signals. 
We have proposed an iterative alternating optimization algorithm based on the matrix FP to tackle the problem of jointly optimizing the downlink precoding, artificial noise covariance, and amplification coefficients  with constraints on transmit powers and information leakage.
Via numerical results, we have confirmed the effectiveness of the proposed scheme compared to baseline schemes.

\end{document}